\documentclass[aps,pre,twocolumn,nofootinbib,floatfix]{revtex4}
\usepackage{amsmath}
\usepackage{epsfig}

\input{tcilatex}

\begin{document}

\title{Entanglement of two coupled molecules}
\author{Y. Y. Liao}
\author{Y. N. Chen}
\author{D. S. Chuu}
\altaffiliation{Corresponding author}
\email{dschuu@mail.nctu.edu.tw}
\affiliation{Department of Electrophysics, National Chiao-Tung University, Hsinchu 300,
Taiwan}
\date{\today }

\begin{abstract}
Entangled states of two coupled polar molecules controlled by laser pulses
are studied theoretically. Schmidt decomposition is used to measure the
degree of entanglement. By varying the pulse shape of the applied laser,
transition from regular to chaotic behavior may occur. This quantum feature
is due to the inference of different phases. Moreover, the entanglement is
also found to be enhanced by increasing the strength of the laser pulses.

\end{abstract}

\maketitle

\address{Department of Electrophysics, National Chiao Tung University,
Hsinchu 300, Taiwan}





The control of the molecular alignment and orientation \cite{1} may be
important in stereodynamics, trapping molecules, molecular focusing, surface
catalysis, and nanoscale design. The alignment is due to the anisotropic
polarizability induced by nonresonant pulses. The pendular states-- hybrid
of field-free molecular eigenstates \cite{2,3,4} --can be created by turning
on a picosecond laser pulse adiabatically. For polar molecules, a
femtosecond laser pulse is able to generate a field-free orientation \cite%
{5,6,7}. \ The molecule remains oriented after turn off the pulse.
Experimentally, several methods concerning the femtosecond photodissociation
spectroscopy and the ion imaging have been employed to measure the rotation
of molecules \cite{8,9,10,11}.

Recently, many researches have been focused to generate entanglement in
quantum-optic and atomic systems \cite{14,15} because entangled states are
fundamental in quantum information processing \cite{12,13}. For molecular
systems, most works are focused on the implementations of quantum
computation \cite{16}. In general, the qubits are encoded by the vibrational
(rotational) modes of the molecules \cite{17}. In order to realize the
quantum logic gates experimentally, the optimal control theory is usually
applied to design the suitable laser pulses. Recently, coupled-rotor- model
attracts much interest also \cite{18} because some physical properties such
as dielectric response may display peculiar behaviors in the presence of
dipole-dipole interaction \cite{19}. However, studies on the entanglement of
coupled molecules under strong laser fields are still lack until now \cite%
{20}. In this rapid communication, a novel method to create entanglement
between two coupled identical polar molecules separated in a distance of
tens of nanometers is proposed. Both molecules are supposed to be irradiated
simultaneously by the ultra-short laser pulses. It is found that the
entanglement induced by the dipole interaction can be enhanced by
controlling the inter-molecule distance or the field strength of the laser
pulse. Moreover, the asymmetrical laser shape is also found to influence
strongly the degree of the entanglement.

Consider now two identical polar molecules separated in a distance of $R$.
The molecule system is irradiated by a series of ultrashort half-cycle laser
pulses \cite{21}. The Hamiltonian of the system can be written as%
\begin{equation}
H=\sum_{j=1,2}\frac{\hbar ^{2}}{2I}L_{j}^{2}+U_{dip}+H_{f-m},
\end{equation}%
where $L_{j}^{2}$ and\ $\frac{\hbar ^{2}}{2I}$ $\left( =B\right) $ are the
angular momentum operator and rotational constant, respectively. $U_{dip}$
is the dipole interaction between two molecules: 
\begin{equation}
U_{dip}=\frac{[\vec{\mu}_{1}\cdot \vec{\mu}_{2}-3\left( \vec{\mu}_{1}\cdot 
\widehat{e}_{R}\right) \left( \vec{\mu}_{2}\cdot \widehat{e}_{R}\right) ]}{%
R^{3}},
\end{equation}%
where $\vec{\mu}_{1}$ and $\vec{\mu}_{2}$ are the dipole moments.\ The
dipole moments of two molecules are assumed, for simplicity, to be
identical, i.e. $\mu _{1}=\mu _{2}=\mu $. The field-molecule coupling $%
H_{f-m}$ can thus be expressed as%
\begin{equation}
H_{f-m}=-\sum_{i=1,2}\mu E\left( t\right) \cos \theta _{i}\cos \left( \omega
t\right) ,
\end{equation}%
where $\theta _{1}$\ and $\theta _{2}$\ are angles between dipole moments
and laser field. The laser profile is assumed in Gaussian shape, i.e. $%
E\left( t\right) =E_{0}e^{-\frac{(t-to)^{2}}{\sigma ^{2}}},$ where $E_{0}$
is the field strength, $t_{0}$ is the center of peak, and $\sigma $ is the
pulse duration. The time-dependent Schr\"{o}dinger equation can be solved by
expanding the wave function $\Psi $ in terms of a series of field-free
spherical harmonic\ functions $Y_{lm}\left( \theta ,\phi \right) $ as 
\begin{equation}
\Psi =\sum_{l_{1}m_{1}l_{2}m_{2}}c_{l_{1}m_{1}l_{2}m_{2}}\left( t\right)
Y_{l_{1}m_{1}}\left( \theta _{1},\phi _{1}\right) Y_{l_{2}m_{2}}\left(
\theta _{2},\phi _{2}\right) ,
\end{equation}%
where $\left( \theta _{1},\phi _{1}\right) $ and $\left( \theta _{2},\phi
_{2}\right) $ are the coordinates of first and second molecule respectively.
The time-dependent coefficients $c_{l_{1}m_{1}l_{2}m_{2}}\left( t\right) $
corresponding to the quantum numbers $\left( l_{1},m_{1};l_{2},m_{2}\right) $
and can be determined by solving the Schr\"{o}dinger equations numerically.
In the above equation, the inter-molecule separation $R$ \ is assumed to be
fixed so that the total wavefunction has no spatial dependence. The
variation of $R$ might be inevitable due to the influence of laser fields or
inter-molecule vibrations. However, recent experiments exhibited that the
spacial resolution in tens of nanometers for two individual molecules
hindered on a surface is actually possible \cite{22,23}. \ The free
orientation model can be easily generalized to hindered ones, in principle,
by replacing the spherical harmonic\ functions with hindered wavefunctions
as shown in our previous work \cite{24}. Thus, the essential physics
discussed here should be realistic.

We now focus our attention on the entanglement generated in this system. The
coupled molecules can be expressed as a pure bipartite system as $\left|
\Psi \right\rangle
=\sum_{l_{1}m_{1}l_{2}m_{2}}c_{l_{1}m_{1}l_{2}m_{2}}\left( t\right) \left|
Y_{l_{1}m_{1}}\right\rangle \left| Y_{l_{2}m_{2}}\right\rangle $. The \emph{%
reduced }density operator for the first molecule is defined as 
\begin{equation}
\rho _{\text{mol 1}}=\text{Tr}_{\text{mol 2}}\left| \Psi \right\rangle
\left\langle \Psi \right| .
\end{equation}%
Following the procedure of Schmidt decomposition, the bases of molecule 1 is
rotated to make the reduced density matrix $\rho _{\text{mol 1}}$ to be
diagonal. The entangled state can be represented by a biorthogonal
expression with positive real coefficients as 
\begin{equation}
\left| \Psi \right\rangle =\sum_{lm}\sqrt{\lambda _{lm}}\left|
Y_{lm}\right\rangle _{\text{mol 1}}\left| Y_{lm}\right\rangle _{\text{mol 2}%
},
\end{equation}%
where $\lambda _{lm}$ is the eigenvalue corresponding to $\left|
Y_{lm}\right\rangle _{\text{mol 1}}\left| Y_{lm}\right\rangle _{\text{mol 2}%
} $. The measure of entanglement for the coupled molecules can be
parametrized by von Neumann entropy 
\begin{equation}
\text{Entropy}=-\sum\limits_{lm}\lambda _{lm}\text{log}_{\text{n}}\lambda
_{lm}.
\end{equation}%
The parameters for numerical calculations are based on NaI molecule whose
dipole moment is 9.2 debyes and rotational constant is 0.12 cm$^{-1}$ in the
ground state. The duration and frequency of the half-cycle pulse are set to
279 fs and 30 cm$^{-1}$, respectively. The center of the peak is 1200 fs and
the initial condition is set as $c_{0000}\left( t=0\right) =1$. The ratio of
the magnitude of positive peak and that of negative peak of the pulse is $%
5:1 $.\ Fig. 1 shows the time-dependent entropy after one pulse passes
through this system. For inter-distance $R=5$ $\times $ $10^{-8}$ m, the
entropy increases slowly from zero to almost finite value of 0.4. For $R=1.5$
$\times $ $10^{-8}$ m, on the contrary, the entropy grows rapidly with the
increasing of time because the dipole force is stronger. Notes that the
entropy only varies within a finite range at long time regime. This
indicates that the systems reaches a dynamic equilibrium state even though
the dipole force is still present. In the insets, the orientations $%
\left\langle \cos \theta _{1}\right\rangle $ and $\left\langle \cos \theta
_{2}\right\rangle $, which can be evaluated immediately after the
coefficients $c_{l_{1}m_{1}l_{2}m_{2}}\left( t\right) $ are determined, also
reflect the effect of the dipole interaction. For $R=5\times 10^{-8}$ m, the
behavior of both molecules is quite close to that of a free rotor \cite{6},
in which the rotational period is close to $\pi \hbar /B.$ As the two
molecules get close enough (the inset of Fig. 1(b)), both molecules orient
disorderly, and the periodic behavior disappears. This is because the dipole
interaction is increased as the distance between the molecules is decreased,
and the energy exchange between two molecules becomes more frequently. The
regular orientation caused by the laser pulse is inhibited by the mutual
interaction.

\begin{figure}[th]
\includegraphics[width=7.5cm]{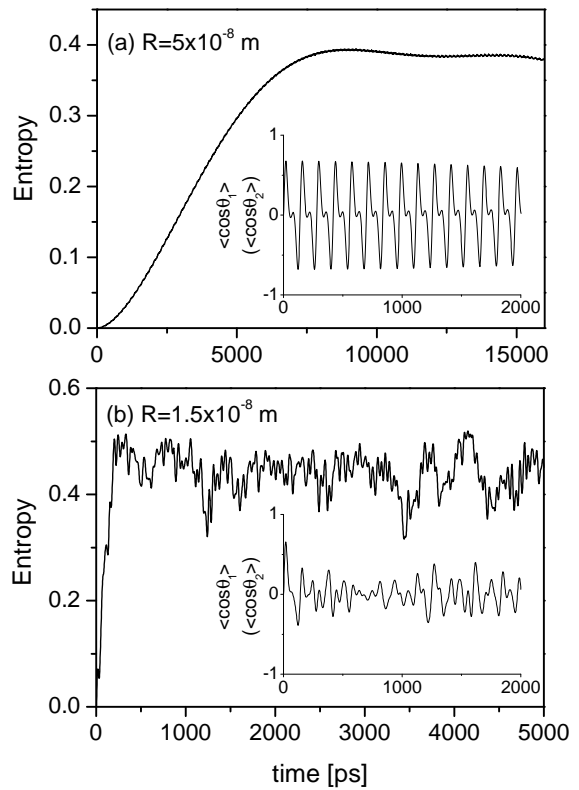}
\caption{Time evolution of the entropy after applying single laser pulse for
(a) $R=5$ $\times $ $10^{-8}$ m and (b) $R=1.5$ $\times $ $10^{-8}$ m. The
insets in (a) and (b) represent the orientations of two molecules at
different distances. The field strength is set equal to $3\times 10^{7}$
V/m. }
\end{figure}

Fig. 2 illustrates the time evolution of the entropy with different ratios
in magnitude of the positive and negative peak value of the laser pulse as $R
$ is set equal to $1.5$ $\times $ $10^{-8}$ m. For the case of ratio $9:1$,
an irregular-like behavior is still obtained, but its time-averaged value is 
$0.51$, which is larger than the averaged one ($0.43$) in Fig. 1. If the
ratio is set equal to $1:1$, the entanglement shows a nearly periodic
behavior with small averaged entropy. This result is very similar to the
case of no laser limit. This implies that the entanglement depends
sensitively on the shape of the laser pulse. The positive and negative parts
of the laser pulse seem to interfere with each other. Moreover, the dipole
force only establishes a periodic-like entropy. 
\begin{figure}[th]
\includegraphics[width=7.5cm]{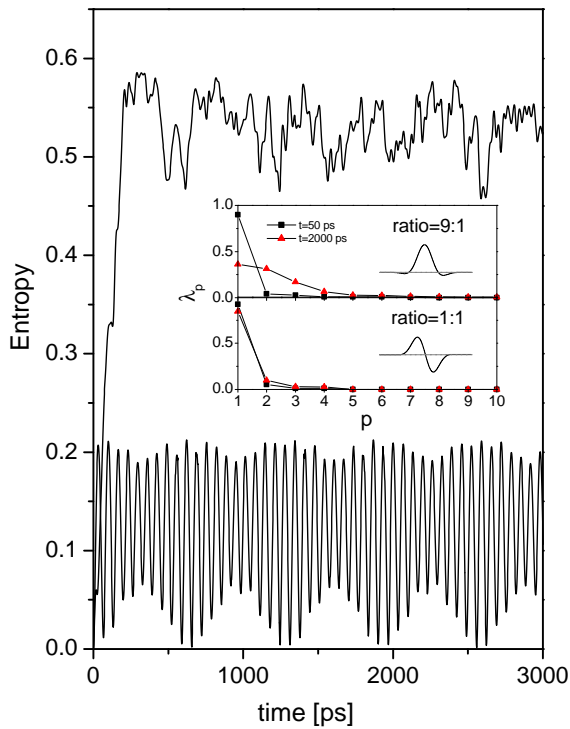}
\caption{Time evolution of the entropy after applying single laser pulse for
different ratios in magnitudes of the positive and negative peak value of
the laser pulse. The graphs show the irregular (periodic) behavior for ratio 
$9:1$ $(1:1)$. The inset : the first ten contributive eigenvalues $\protect%
\lambda _{\text{p}}$ at short time $(t=50$ ps$)$ and long time $(t=2000$ ps$)
$. The field strength and inter-molecule separation are $E_{0}=$ $3\times
10^{7}$ V/m and $R=1.5$ $\times $ $10^{-8}$ m.}
\end{figure}

The distributions of the first ten contributive eigenvalues ($\lambda _{%
\text{p}})$ to the entropy at short and long time regimes are shown in the
insets of Fig. 2. In the case of $9:1$ ratio, the eigenvalue $\lambda _{1}$
dominates the contributions at short time regime($t=50$ ps). However, the
dominant contributions are distributed to several values as $t=2000$ ps
regime. This means in long time limit the system is in some sort of dynamic
equilibrium, and entropy saturates to certain value. On the contrary, $%
\lambda _{1}$ always dominates the contributions for either short or long
time regime in the case of symmetrical laser shape as shown on the lower
inset of Fig. 2. From statistical point of view, this somehow explains the
suppressed and regular behaviors of the entanglement (entropy).

Fig. 3 shows the time evolutions of the populations of the eigenstates for
different ratio of pulse shapes. For $1:1$ ratio, the energy level $%
(0,0;0,0) $ is mostly populated as shown in the lower panel of Fig. 3. Like
the ground state, the populations of the higher levels (the inset of Fig. 3)
also show the periodic behavior. The reason is attributed to the symmetrical
shape of the laser so that the populations of the higher levels are almost
contributed by the dipole interaction. In this case, the magnitudes of the
higher level populations are rather small such that the periodic evolution
of the entropy is obtained. On the other hand, for $9:1$ ratio the
populations of the higher states represent different degrees of irregularity
as shown in the upper panel of Fig. 3. This is because a single laser pulse
can generate high populations in the excited states \cite{6}. Energy
transfer by means of (mediated) dipole interaction generates the irregular
evolutions of the higher excited states, which result in a randomly
time-varying entropy. 
\begin{figure}[th]
\includegraphics[width=7.5cm]{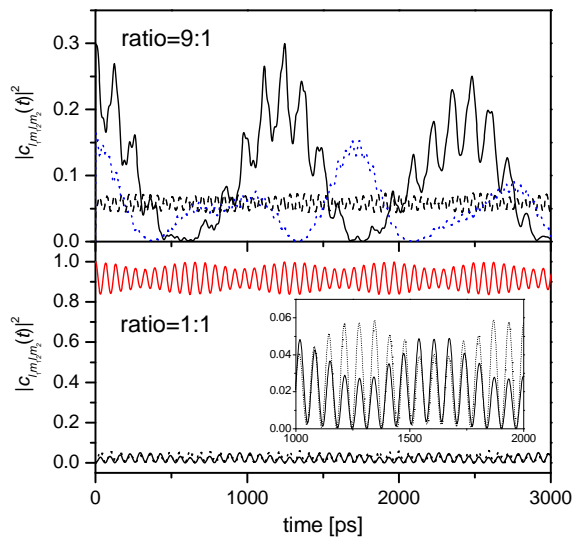}
\caption{Populations of the states $\left( l_{1},m_{1};l_{2},m_{2}\right) $
for different ratios. Upper panel : $\left( 1,0;0,0\right) $ (dashed line), $%
\left( 1,0;1,0\right) $ (solid line), $\left( 2,0;1,0\right) $ (blue dotted
line). Lower panel : $\left( 0,0;0,0\right) $ (red line), $\left(
1,0;1,0\right) $ (solid line), $\left( 1,1;1,1\right) $ (dotted line). The
inset in the lower panel is the enlarged figure showing the states $\left(
1,0;1,0\right) $ (solid line), $\left( 1,1;1,1\right) $ (dotted line),
respectively.}
\end{figure}

By adjusting the laser parameters, one can vary the degrees of the
entanglement. Fig. 4 illustrates the time evolution of the entropy for fixed
inter-molecule separation $R=1.5$ $\times $ $10^{-8}$ m and laser ratio $5:1$
for a doubling of the laser field strength ($E_{0}=6\times 10^{7}$ V/m). As
can be seen, an irregular-like behavior of the entropy is obtained, and its
averaged value is larger in comparison with Fig. 1(b). This can be
understood well by studying the relationship between the dipolar interaction
and the field strength. If the effect of laser field overwhelms the dipole
interaction, the populations are distributed to a wider range. Since most of
the populations are also distributed more averagely in this case, the
entropy from Schmidt decomposition is certainly larger. Another way to
control the degree of entanglement in this system is to change the positive
and negative ratios of the laser pulse. Inset of Fig. 4 shows the \emph{%
time-averaged} entropy with respect to different ratios. We find that the
entropy is more enhanced as the ratio is larger. This means that the high
asymmetric shape of laser pulse can generate a larger entropy at the same
field strength. 
\begin{figure}[th]
\includegraphics[width=7.5cm]{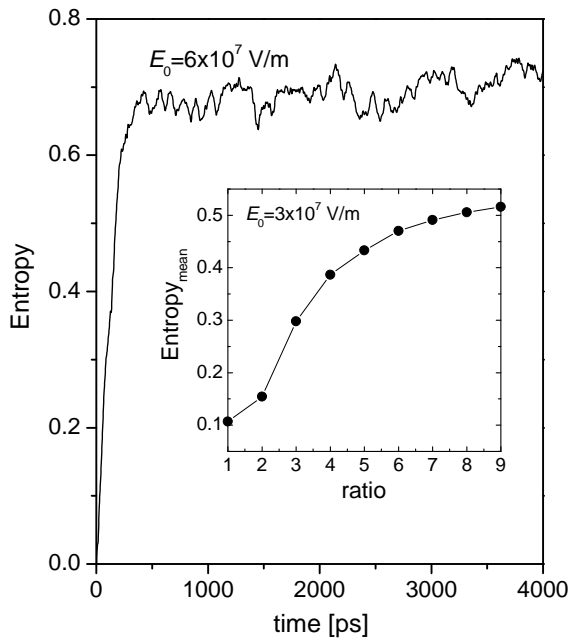}
\caption{Time evolution of the entropy for fixed field strength $%
E_{0}=6\times 10^{7}$ V/m and ratio $5:1$. The inset : Dependence of the
time-averaged entropy on the pulse shape for fixed field strength $%
E_{0}=3\times 10^{7}$ V/m and inter-molecule separation $R=1.5$ $\times $ $%
10^{-8}$ m}
\end{figure}

A few remarks about the differences between our model and previous works on
generating entangled states should be addressed here. In our model, we
consider the rotational excited states instead of internal \emph{electronic}
states of the molecules. Second, the laser frequency in our work is tuned as
possible as far-away from resonance, while conventional creation of
entanglement depends on the resonant driving pulses. This means our work
provides a wider range to select the laser frequency to generate
entanglement. As for the effect of decoherence, our entanglement is formed
by the excited rotational states, instead of the vibrational states,
therefore, the decoherence is dominated by photon emission even if the
molecules are attached to the surface of a solid.

In conclusion, we have studied the entanglement of two coupled polar
molecules irradiated by an ultra-short laser pulse. Schmidt decomposition is
used to generated the time evolution of the entropy. The degree of the
entanglement is analyzed by varying the inter-molecule distance, symmetry of
the laser shape, and field strength. Its dependence on the controlled
parameters may be useful for the quantum information processing.

We would like to thank to C. M. Li for helpful discussions. This work is
supported by the National Science Council, Taiwan under the grant numbers
NSC 92-2120-M-009-010 and NSC 93-2112-M-009-037.

\end{document}